\documentclass{osa-article}

\journal{oe}

\usepackage{subfigure}
\usepackage{graphicx}
\usepackage{dcolumn}
\usepackage{bm}
\usepackage{subfigure}
\usepackage{graphicx,epsfig,epstopdf}
\usepackage{amsmath}
\usepackage{color}
\usepackage{subfigure}
\usepackage{graphicx}
\usepackage{dcolumn}
\usepackage{bm}
\usepackage{subfigure}
\usepackage{graphicx,epsfig,epstopdf}
\usepackage{amsmath}
\usepackage{color}
\usepackage{subfigure}
\usepackage{graphicx}
\usepackage{dcolumn}
\usepackage{bm}
\usepackage{subfigure}
\usepackage{graphicx,epsfig,epstopdf}
\usepackage{amsmath}
\usepackage{color}
\usepackage{subfigure}
\usepackage{graphicx}
\usepackage{dcolumn}
\usepackage{bm}
\usepackage{subfigure}
\usepackage{graphicx,epsfig,epstopdf}
\usepackage{amsmath}
\usepackage{color}
\usepackage{subfigure}
\usepackage{graphicx}
\usepackage{dcolumn}
\usepackage{bm}
\usepackage{subfigure}
\usepackage{graphicx,epsfig,epstopdf}
\usepackage{amsmath}
\usepackage{color}
\usepackage{subfigure}
\usepackage{graphicx}
\usepackage{dcolumn}
\usepackage{bm}
\usepackage{subfigure}
\usepackage{graphicx,epsfig,epstopdf}
\usepackage{amsmath}
\usepackage{color}
\articletype{Research Article}

\begin{document}

\title{\centering{Self-biased Tri-state Power-Multiplexed Digital Metasurface Operating at Microwave Frequencies}}

\author{Mehdi Kiani\authormark{1}, Majid Tayarani\authormark{1,*}, Ali Momeni\authormark{2}, Hamid Rajabalipanah\authormark{2}, and Ali Abdolali\authormark{2}}

\address{\authormark{1}Department of Electrical Engineering, Iran University of Science and Technology, Tehran 1684613114, Iran\\
\authormark{2}Applied Electromagnetic Laboratory, School of Electrical Engineering, Iran University of Science and Technology, Tehran 1684613114, Iran
\\
}

\email{\authormark{*}m\_tayarani@iust.ac.ir} 



\begin{abstract}
	Exploiting of nonlinearity has opened doors into undiscovered areas for achieving multiplexed performances in recent years. Although efforts have been made to obtain diverse nonlinear architectures at visible frequencies, the room is still free for incorporating non-linearity into the design of microwave metasurfaces. In this paper, a passive dual-band power intensity-dependent metasurface is presented. Here, power-multiplexing performance is achieved by embedding PIN-diodes in the coding particles. The proposed digital metasurface has three operational states: 1) it acts as a normal reflector at low power intensities whereas provides a dual-band nonlinear response upon illuminating by high-power incidences where 2) it perfectly absorbs the radiations at $f_1$=6.7 GHz and 3) re-distributes the scattered beams by arranging the meta-atoms with a certain coding pattern at $f_2$=9.37 GHz. The nonlinear performance of the designed tri-state coding elements has been characterized by using the scattering parameters captured in the full-wave simulations where the accurate models of the nonlinear diodes are involved. The emergence of microwave self-biased metasurfaces with smart re-actions upon illuminating by incidences of different power level, revealing great opportunities  for designing smart windows, smart camouflage coating surfaces and so on.
\end{abstract}

%
%
%
%
%

\section{Introduction}
\begin{figure*}[t]
	\centering
	\includegraphics[height=4in]{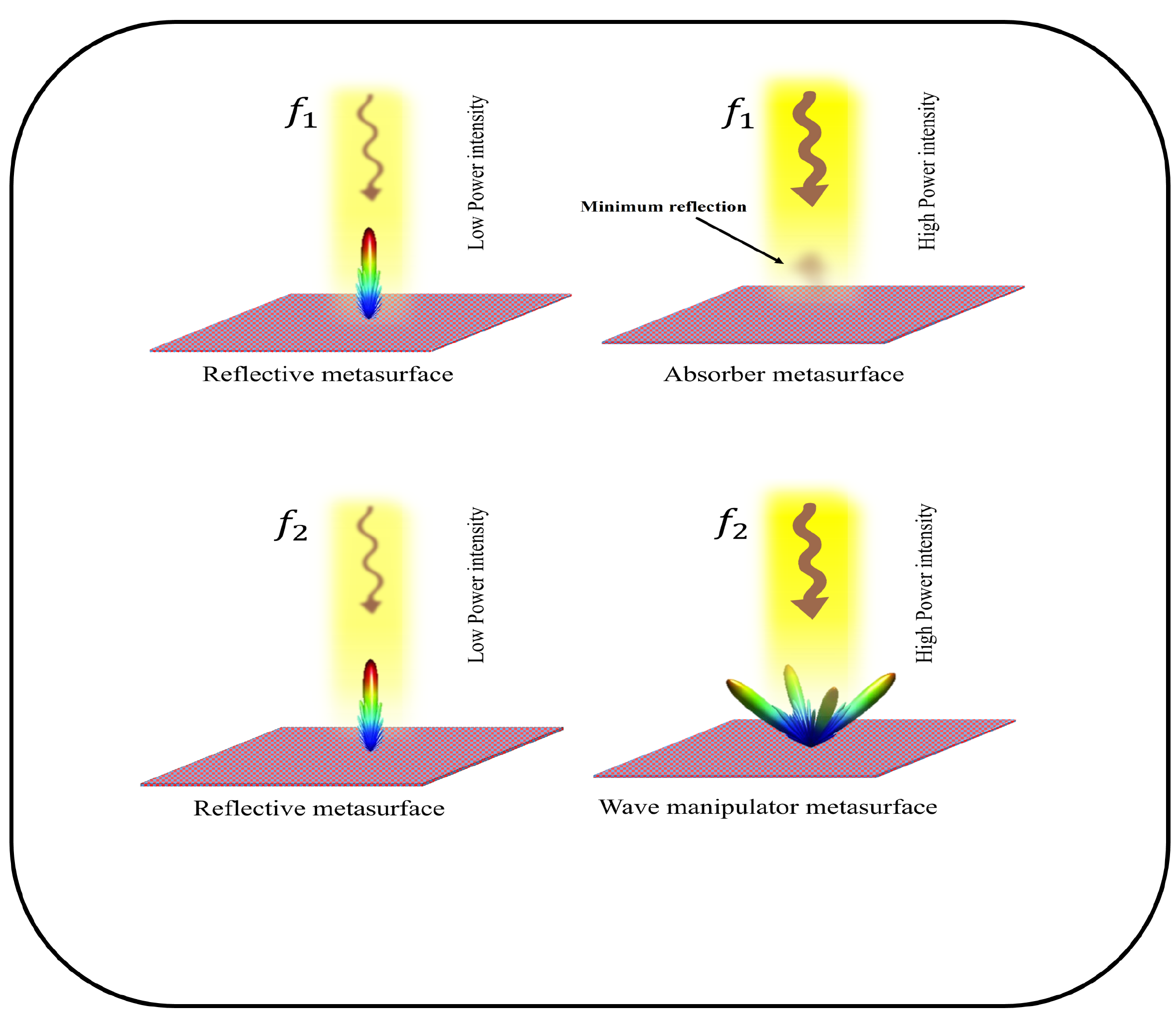}
	\caption{\label{fig:epsart} {A schematic model of the smart reflective metasurface. The designed metasurface acts as an EM mirror upon illuminating by low-power radiations. While the intensity is increased, the nonlinearity plays dominant role and the metasurface turns into a bi-spectral low-scattering surface exposing microwave absorption at ($f_1$) and scattering diffusion at ($f_2$). } }
\end{figure*} 
~~~Metasurfaces as the reduced dimensional version of 3D metamaterials can have the same electromagnetic (EM) functionalities as many bulky components while benefiting from extraordinary features of being ultra-thin, easy to build, and easy to integrate. These unique properties have received a broad interest in the engineering, physics and optics communities, leading to extensive theoretical and experimental investigations. Since their creature, a vast variety of intriguing potential applications has been reported for metasurfaces from microwave to visible spectra. The field of metasurface has witnessed  rapid growth, and found a special place in obtaining negative refraction \cite{shelby2001experimental}, polarizers \cite{zhao2011manipulating}, anomalous reflectors \cite{sun2012high}, engineered antennas \cite{lin2014dielectric}, beam scanning \cite{moeini2019wide,moeini2019collimating}, invisibility cloak \cite{pendry2006controlling,li2008hiding}, and advanced analog computing \cite{momeni2019generalized,abdolali2019parallel,momeni2019tunable} . With the advent of generalized Snell's laws of reflection and refraction \cite{yu2011light,rajabalipanah2019asymmetric}, the demands for any types of controllable  metasurfaces have been gradually increased for which phase change materials, graphene \cite{fallahi2012design,cheng2015dynamically,rahmanzadeh2018multilayer}, vanadium dioxide \cite{liu2016hybrid,cueff2015dynamic,hashemi2016electronically} and semiconductors \cite{ratni2018active,li2017electromagnetic} manifest themselves as reliable solutions. 

In recent years, a copious number of studies has incorporated nonlinearity into the metasurface design, yielding different multitasking types like frequency multiplexing \cite{ye2016spin}, harmonic generation \cite{rose2011controlling,li2015continuous}, and so forth. In the microwave regime, several proposals have elaborately explored power-dependent metasurfaces by embedding PIN-diodes into the building unit cells. PIN-diodes, while the power intensity is more than a specified threshold, are in On-state, acting as a small resistor; conversely, when the power intensity is less than that threshold, the diode operates in Off-state, acting as a small capacitor. In 2013 \cite{wakatsuchi2013circuit,wakatsuchi2013waveform}, through the use of diodes integrated into the metasurface, the Sievenpiper's group introduced the concept of circuit-based nonlinear absorbers that absorb high power surface currents but not small signals. However, the proposed structure acts only upon exciting by high power surface waves propagating on the nonlinear metasurfaces. In 2019, Zhao \textit{et. al} \cite{zhao2017power} designed a frequency selective surface to provide a passband for low-power microwave signals and a stopband for high-power microwave signals. Nevertheless, the presented architecture can only manipulate the spectral (not spatial) properties of the high power incident signals. 

Digital metasurfaces were originally put forward to engineer the scattered waves, by designing two distinct coding elements with opposite reflection phases (e.g., 0$^\circ$ and 180$^\circ$) as the “0” and “1” digital bits \cite{cui2014coding,momeni2018information,rouhi2018real,rouhi2019multi,hosseininejad2019digital}. Through organizing the coding particles on a 2D plane with pre-determined coding sequences, such architectures can serve to manipulate EM waves in a simple and effective way. They could greatly facilitate the optimization procedures since the phase responses are digitally discretized in the microscopic and macroscopic level designs, thereby noticeably alleviating the parameter search space. More recently, Z. Luo \textit{et. al} \cite{luo2019intensity} proposed an intensity-dependent nonlinear metasurface whereby the phase profile on the surface can be digitally determined by the incoming intensity, opening up new possibilities in nonlinear EM wave manipulations. However, besides resorting into biasing voltages provided by external active circuits, the operating band of the digital metasurface was restricted to single narrow frequency region. 

In this paper, a self-biased dual-band coding metasurface is proposed, comprising of linear and nonlinear power-multiplexed I-shape meta-atoms, as illustrated in \textbf{\textcolor{blue}{Figure 1}}. At low power intensities, the designed metasurface functions linearly and the coding status of all digital particles are the same, resulting in a broadband EM mirror. At high power intensities, the nonlinear property of some meta-atoms smartly turns the linear reflector into a bi-spectral nonlinear coding metasurface with two separate camouflage functionalities. In the lower band (C-band), the nonlinear coding metasurface perfectly absorbs more than 99 $\%$ of the incident wave power while in the higher band (X-band), the disordered coding pattern of the anti-phase linear and nonlinear particles, as 0 and 1 digital bits of the metasurface, dramatically reduces the specular reflection by generating multiple randomly-oriented scattered beams. Indeed, the operational status of the self-biased tri-state (EM mirror, absorber, and diffuser) digital metasurface is merely adjusted by the power intensity of the incident wave, without need to any external biasing circuit. The numerical simulations verify the performance of the proposed bi-functional nonlinear coding metasurface. 
\begin{figure*}[t]%
	\centering
	\subfigure[]{%
		\label{fig:21}%
		\includegraphics[height=2in]{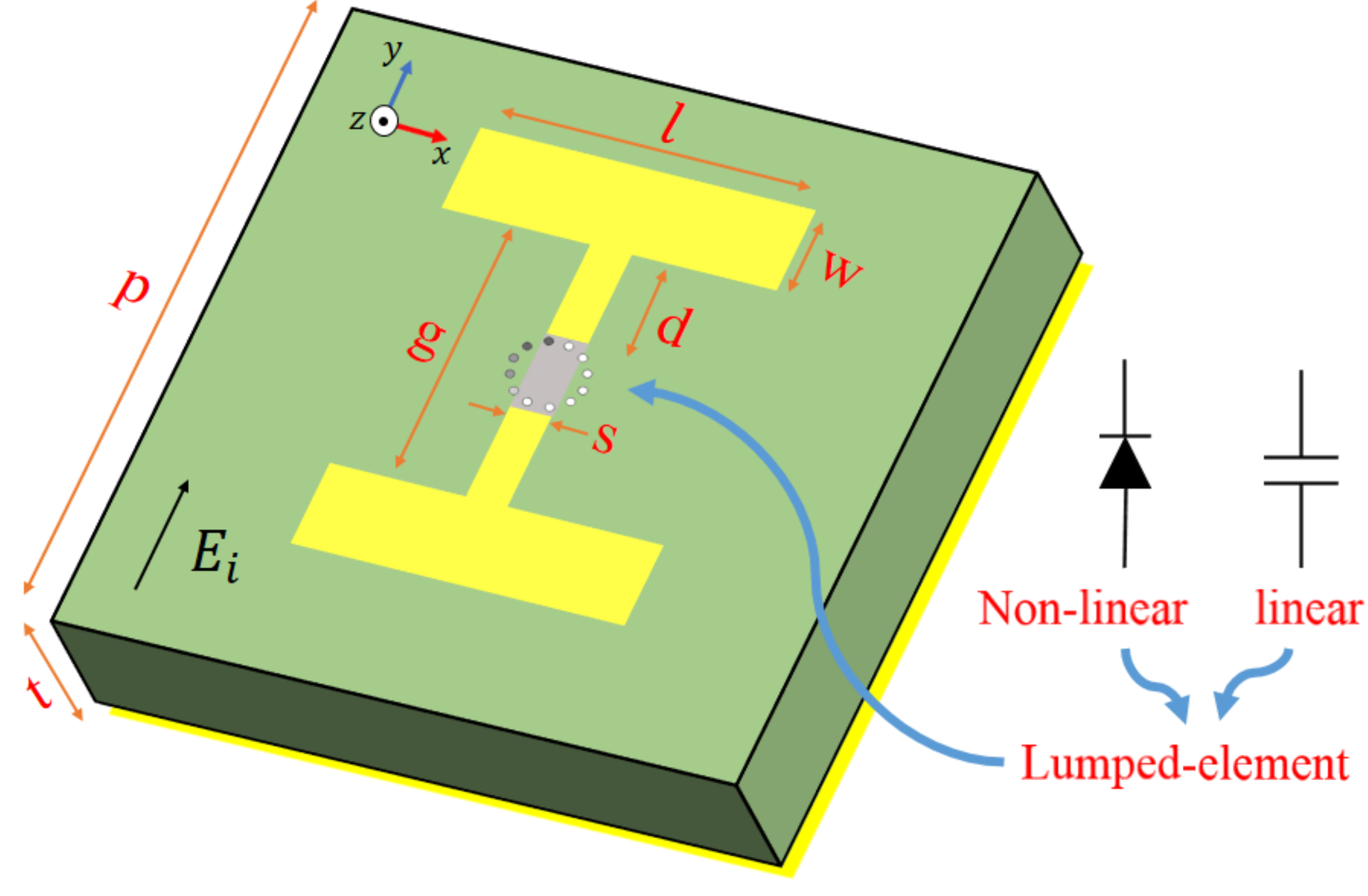}}%
	\subfigure[][]{%
		\label{fig:22}%
		\includegraphics[height=2 in]{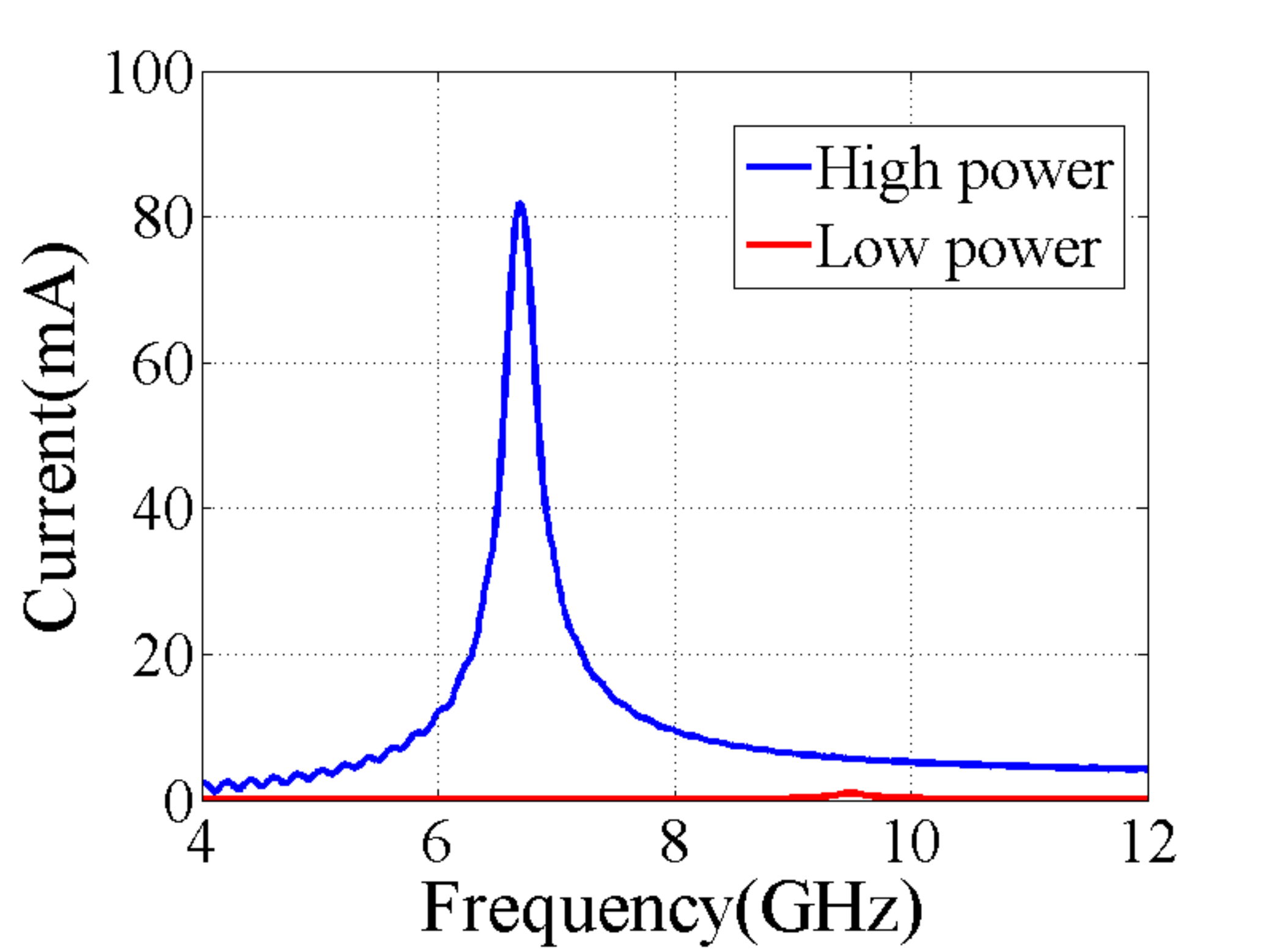}}%
	\caption{ (a) I-shaped unit as the basic meta-atom of the proposed metasurface. When the cell uses a diode, the unit cell is called as  nonlinear meta-atom, but when using the capacitor the unit cell is called as linear meta-atom. (b)The spectra of the current induced on the PIN-diode of the nonlinear meta-atom upon illuminating by low-power and high-power incidences.  }
	\label{fig:4}
\end{figure*}

\section{Coding Meta-atom Design}

\subsection{Principle Mechanism} 
~~The coding metasurface of \textbf{\textcolor{blue}{Figure 1a}} is digitally built by two different linear and nonlinear meta-atoms. Upon illuminating by low-power radiations, the two groups of meta-atoms possess the same digital state (0 or 1) at the linear mode, reflecting the incident wave with identical reflection phase and amplitude. However, when exciting by high-power incidences, the coding status of the nonlinear elements smartly toggles (1 or 0), realizing a certain coding pattern over the surface. The overall geometry of the designed coding meta-atoms is depicted in \textbf{\textcolor{blue}{Figure 2a}}, consisting of three separate layers: 1) the I-shape meta-atom as the top layer which is loaded in the middle gap by a Macom MA4L401-134 PIN-diode \textcolor{red}{ }, 2) the Rogers RT5880 substrate with the thickness of $h$=1 mm as the middle layer, and 3) a copper ($\sigma=5.96 \times 10^7 $) ground plane as the bottom layer avoiding the energy transmission across the frequency band of study. They are all made by printed circuit board (PCB) technology. The diode can be circuitally modeled as $R$=1.2 and $C$= 0.2 pF in the nonlinear working region while it can be represented by C=0.2 pF (Murata-GJM1555C1HR20WB01D) in the linear operating mode \textcolor{red}{.} The structural parameters of both linear and nonlinear particles can be listed as $l$=6 mm, $p$=11 mm, and $t$=1 mm. However, $w_{lin}$=3.2 mm, $s_{lin}$=0.4 mm $g_{lin}$=3.88 mm, and $d_{lin}$=1.47 mm while $w_{nonlin}$=2 mm, $s_{nonlin}$=3.8 mm, $g_{nonlin}$=3 mm, and $d_{nonlin}$=1.31 mm for linear and nonlinear groups of meta-atoms, respectively. 

The principle mechanism of the nonlinear group of coding particles is based on changing the value of the equivalent lumped elements upon increasing the power level of the incident wave. Obviously, the voltage levels induced on both ends of the embedded diode strongly depend on the local field intensities stimulated by different incident power levels. If the induced voltage is not enough to turn the diode ON, it can be modeled as a small capacitance. In contrast, at the condition of high power illumination, such AC voltage can bias the diode mounted in the gap positively or negatively. For PIN diodes, carriers are injected into intrinsic layer during the half-cycle of positive bias and migrated out of intrinsic layer during the half-cycle of negative bias. Because
the carrier injection rate is higher than the outgoing rate, a certain charge will be accumulated in the
intrinsic layer, and the diode will work stably in the ON-state.

\subsection{Time-domain Analysis}
~~The nonlinearity manifests itself to the time-domain response of the nonlinear group of meta-atoms, clearly. In this section, a Macom MA4L401-134 PIN-diode with low parasitic capacitance, short switching time, and small ON-state resistance has been chosen. The nonlinear diode acts as a capacitance of about 0.25 pF at OFF-state, while it is equivalent to a series of resistor 1.2 $\Omega$ at ON-state. In our simulations, lumped element capacitors and resistors are deliberately employed to describe both OFF and ON states of the diodes, respectively. The commercial program, CST Microwave Studio, is utilized to observe the time domain waveforms of voltages induced on both terminations of the PIN-diodes, in low- and high-power situations. Periodic boundary conditions are applied to the x- and y-directions,  while Floquet ports are assigned along the z-direction to form a transversally-infinite array from the meta-atoms of \textbf{\textcolor{blue}{Figure 2a}}. The polarization of the incident EM field is set to be parallel with the middle metallic lines of the I-shape coding elements. 
A probe was involved in the simulations to capture the voltage waveforms at both ends of the PIN-diode upon exciting by different power intensities. The voltages of the OFF and ON states are recorded and shown in \textbf{\textcolor{blue}{Figure 3}}, respectively. When the small signal impinges on the coding element, the voltage waveform at both ends of the diode behaves as a complete sinusoidal wave, indicating that such low-power signal cannot turn on the diodes, and thus does not switch the coding status of the I-shape meta-atom. However, shinning the coding element with large signal inputs makes the waveforms at both ends of the diode distorted, demonstrating that the I-shape meta-atom reaches its nonlinear working region with power-dependent EM characteristics. As expected, since the forward voltage of the employed PIN-diode is reported as 0.75 V, the positive half-cycle voltage is successfully truncated at the level of 0.75 V when the diode is triggered by high power inputs. When a high-power plane wave impinges on the nonlinear particles, the PIN-diodes are in ON-state, although there is no any external biasing circuit. In fact, the I-shape meta-atoms are quite passive and adopt themselves with the power level of the input signals, making us call the metasurface as a self-biased architecture. Switching speed of the employed PIN-diodes between ON and OFF states is 10 nS. Since the {minimum} time period of the input waveform, 0.15 nS, is very smaller than the switching speed, the speed of the electrons/holes injected to the P- and N-type junctions in the positive cycle is much more than that of the outgoing electrons/holes during the negative cycle. This confirms that the PIN-diodes are always in ON-state operational mode when they are exposed to the high-power radiations \cite{wu2019energy,zhao2019power}

\subsection{Frequency-domain Analysis}

~~ The overall EM functionality of the designed nonlinear metasurface is determined by the reflection phase/amplitude spectra of the occupying digital  elements and their spatial distribution over the surface. 
In this section, we aim at characterizing the EM responses of linear and nonlinear coding particles in the frequency domain. The EM response of the linear coding particles can be simply characterized using the conventional methods where the PIN-diode is replaced with its equivalent resistance. The corresponding results are given in \textbf{\textcolor{blue}{Figures 4a-c}}. The reflection properties of the linear particles are regardless of the intensity variations. 

The minimum input power changing the status of the nonlinear elements can be specified through monitoring the current induced across the diode for different power levels. The results are \textcolor{black}{ illustrated in  \textbf{\textcolor{blue}{Figure 2b}}.} Under high-power illuminations (e.g. 10 dBm), PIN-diodes function in the ON-state because the induced current exceeds the standard threshold; But, being exposed to low-power plane waves (e.g. $-$10 dBm), the current passing through the PIN-diode of nonlinear meta-atoms is far less than the standard threshold and the diodes are in OFF-state. 
\begin{figure*}[t]
	\centering
	\includegraphics[height=2in]{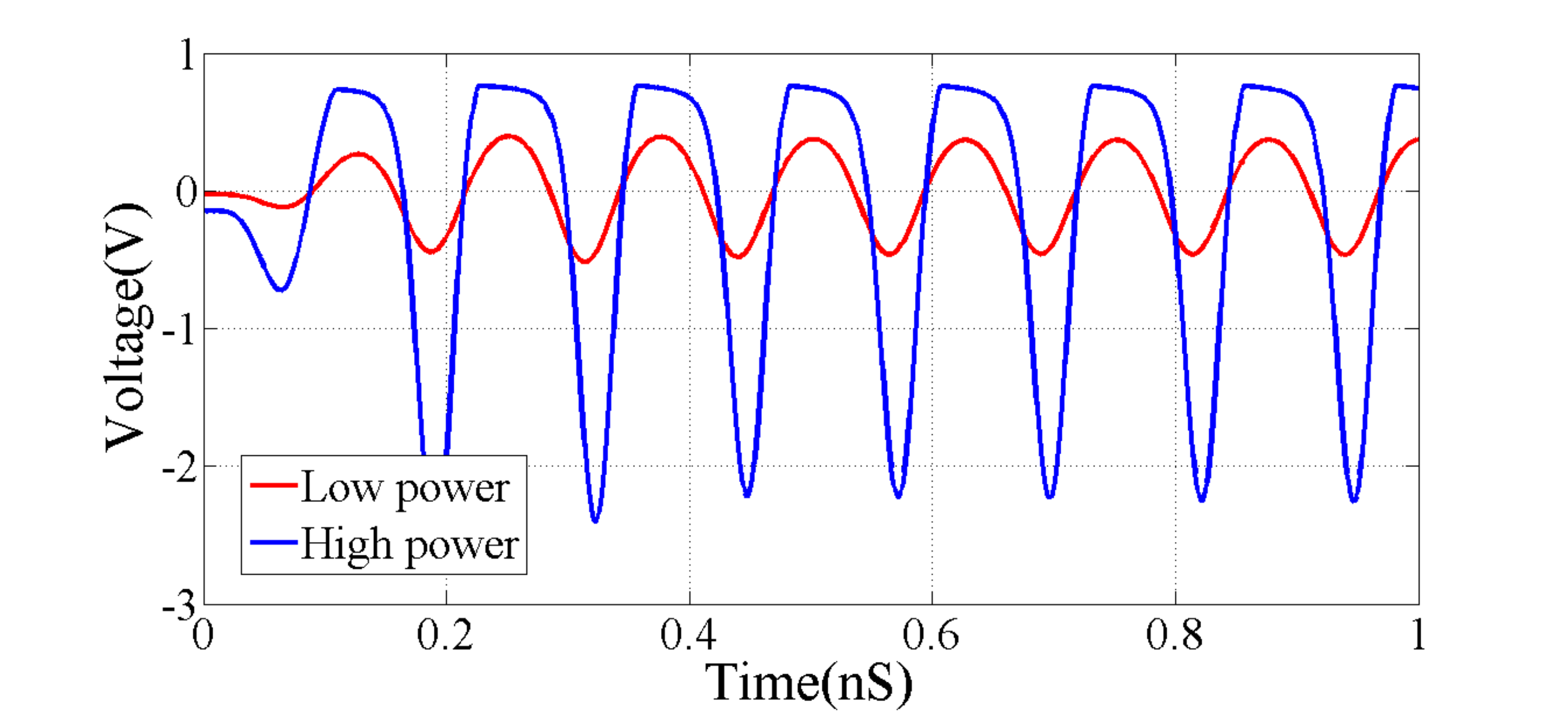}
	\caption{\label{fig:epsart}  { {The time domain response pertaining to the voltage signals induced at both ends of the diode at high and low power intensities .} }}
\end{figure*}

When a y-polarized plane wave excites each particle, the complex reflection coefficient of the nonlinear meta-atoms can be tuned by the voltages induced across the PIN-diode. Using EM-circuit co-simulations carried out by CST Microwave Studio, the dispersive reflectivity of the nonlinear group of elements is studied for two different power levels. In this case, the diode is numerically replaced with a lumped port, allowing us to extract the S2P data. The numerical simulations of the nonlinear elements return the scattering parameters, i.e. the reflection coefficients on the reference plane. The scattering parameter module is then imported to the Circuit Simulator, where the second port is terminated by the SPICE model of the PIN-diode and the first one is connected to an external microwave source. After the circuit simulations, the complex reflection coefficient of the nonlinear particles are recorded and plotted in \textbf{\textcolor{blue}{Figures 4a-c}} for low- and high-power input signals, respectively. The first deduction of comparing \textbf{\textcolor{blue}{Figure 4a}} and \textbf{\textcolor{blue}{Figure 4c}} is that upon illuminating by low-power signals, both linear and nonlinear coding particles mimic the “0” digital state with high reflectivity across 4 GHz to 14 GHz. In this case, the structure mimics the response of an EM mirror.  In contrast, when the nonlinear particles are exposed to high-power plane waves, a bi-spectral EM response is observable. At $f$=6.7 GHz, the array of nonlinear meta-atoms plays the role of a perfect absorber with more than 99.9$\%$ energy dissipation occurring in the equivalent resistance (see \textbf{\textcolor{blue}{Figure 4a}}). Therefore, the metasurface can serve well as an energy harvesting platform in this frequency band. The absorption result for the random arrangement of elements is given here. In the vicinity of $f$=9.3 GHz, however, the coding status of the nonlinear meta-atom is changed to “1”, because of showing a 180$^\circ$ phase difference with respect to the linear particles (see \textbf{\textcolor{blue}{Figure 4b}}). Furthermore, the binary phase response required for constructing the coding metasurface is acquired at the higher frequency band. The reflection magnitude of both linear and nonlinear coding particles is high enough to reveal the high reflectivity required for highly-efficient EM wave manipulation (see \textbf{\textcolor{blue}{Figure 4c}}). The 0.86 magnitude is due to intrinsic resonance of the I-shape meta-atom in the presence of external capacitor.  In summary, the nonlinear intensity-dependent reflection properties of the I-shape inclusions would bring up tri-state functionalities: 1) EM mirror, 2) energy harvester, and 3) coding metasurface within a shared aperture.

\begin{figure*}[t]%
	\centering
	\subfigure[]{%
		\label{fig:21}%
		\includegraphics[height=1.8in]{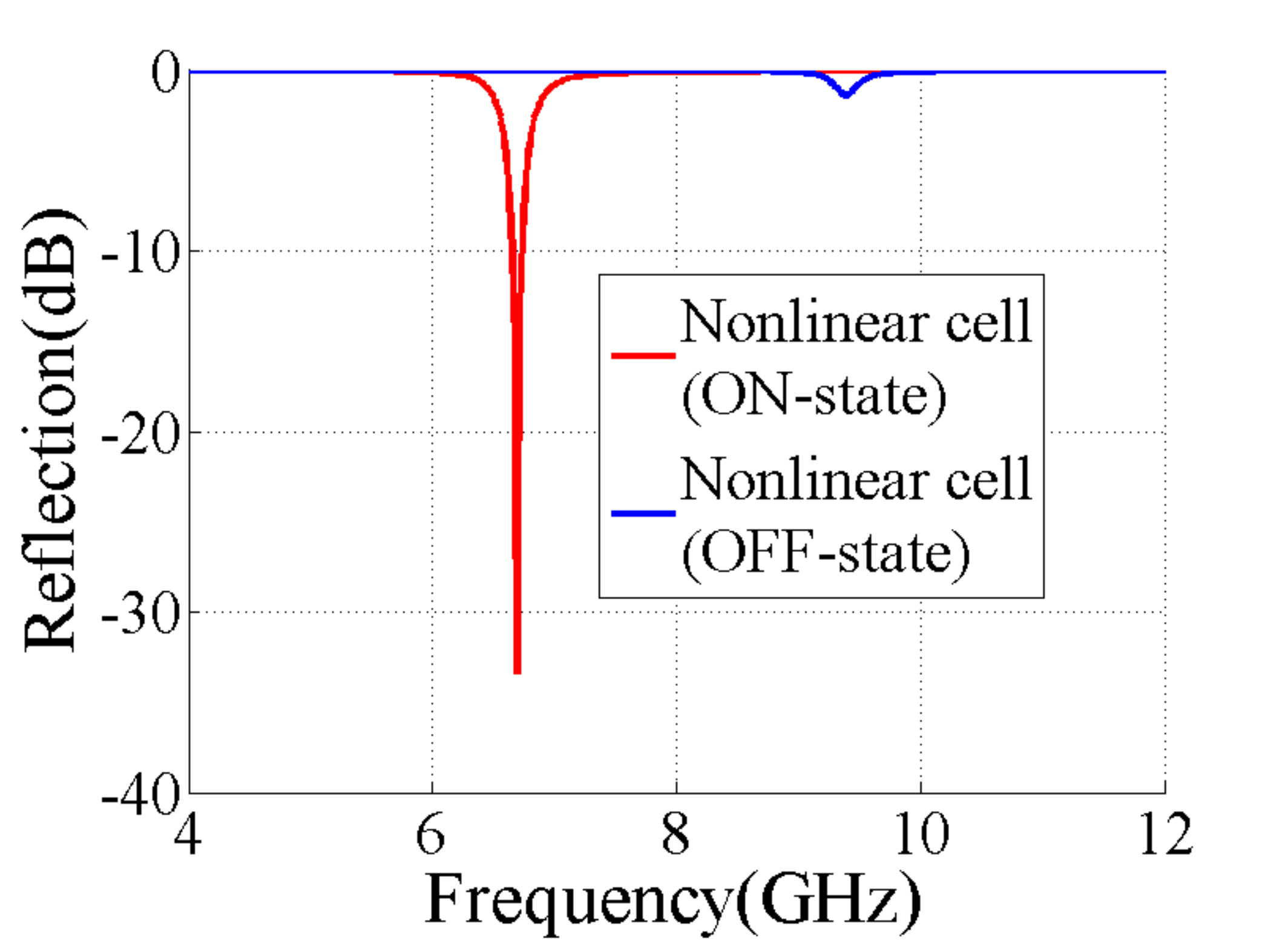}}%
	\qquad
	\subfigure[][]{%
		\label{fig:22}%
		\includegraphics[height=1.8 in]{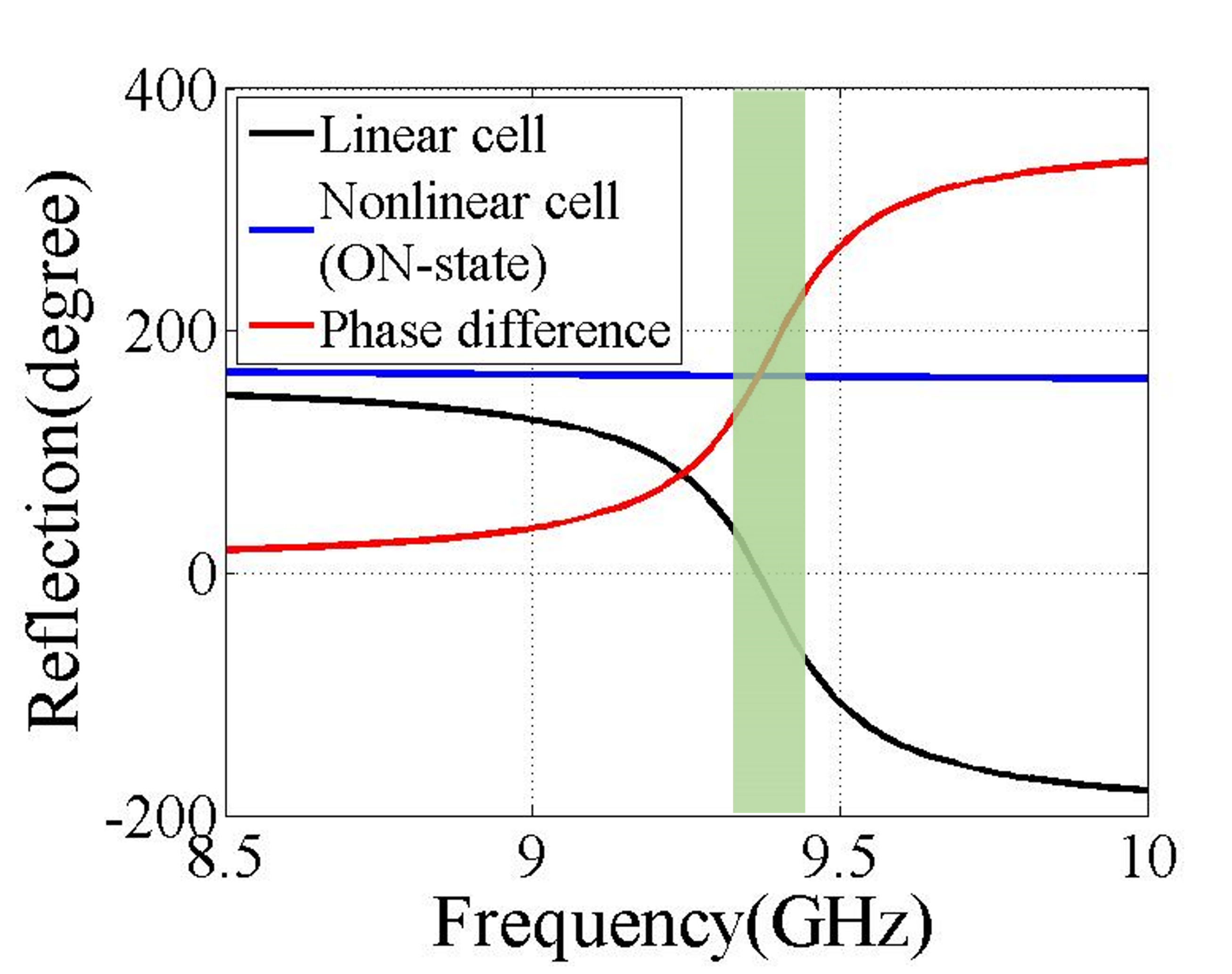}}%
	\subfigure[][]{%
		\label{fig:22}%
		\includegraphics[height=1.8 in]{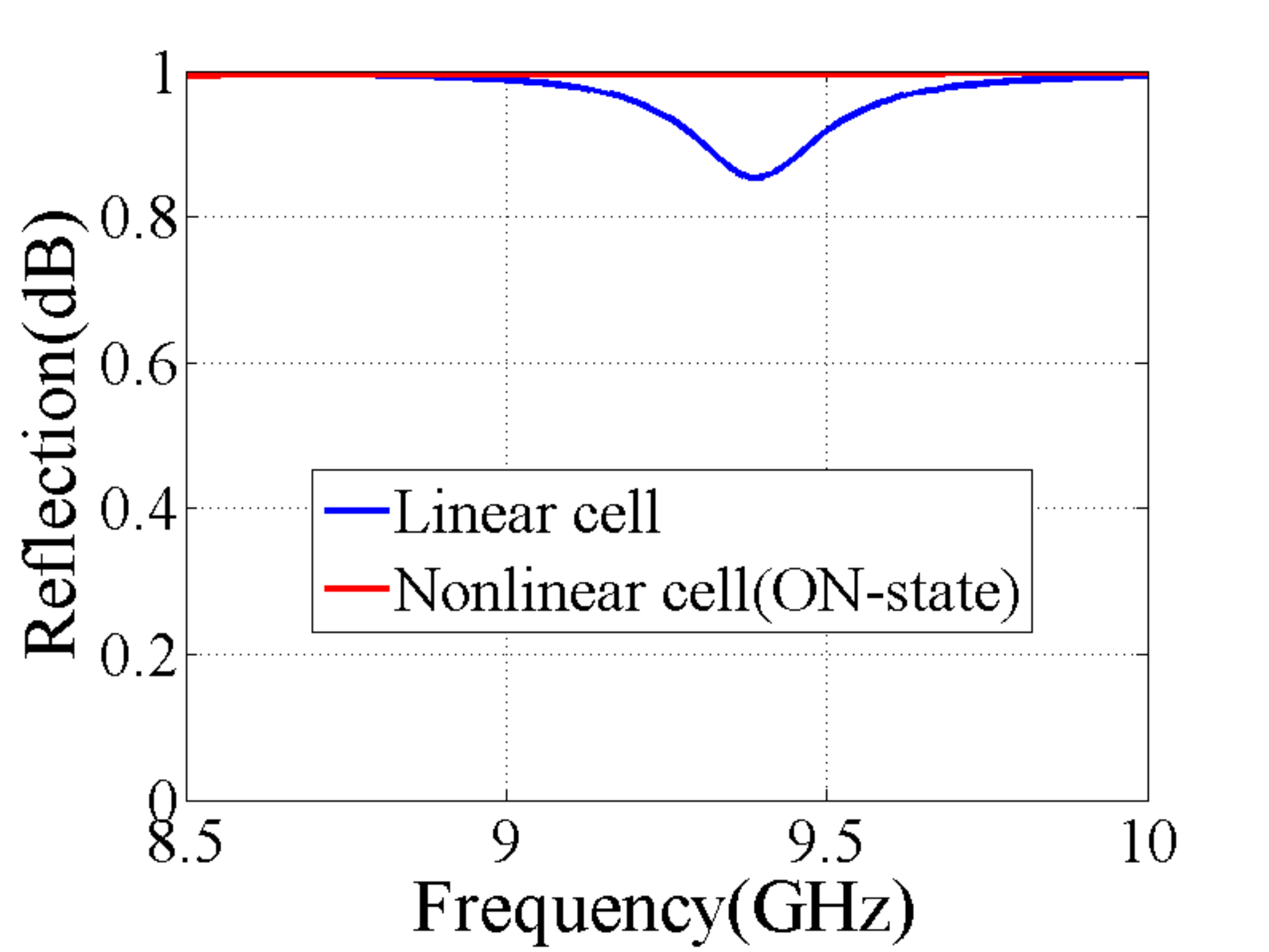}}%
	\caption{ Simulation results of the reflection magnitude of the linear and nonlinear meta-atoms for the low-power (OFF-state) and high-power (ON-state) illuminations at (a) lower and (c) upper frequency bands.  (b) Simulation results of the reflection phase for the linear and nonlinear meta-atoms exposed to low-power and high-power radiations of the upper frequency band.}
	\label{fig:4}
\end{figure*}

\section{Results and Discussion}
~~As depicted in \textbf{\textcolor{blue}{Figures 5}}, the proposed architecture is composed of linear and nonlinear I-shape meta-atoms whose spatial distribution over the surface constructs a tri-state intensity-dependent bi-spectral metasurface. In order to suppress the corner coupling effects, 4 $\times$ 4 linear I-shape meta-atoms are integrated into a lattice. The same scheme is realized for nonlinear lattices. The linear and nonlinear lattices are marked by red and blue colors where different arrangements of nonlinearity would lead to various EM functionalities at the upper frequency band. The nonlinearity distribution can be arbitrary. The tri-state metasurface with different arrangements of nonlinearity has been built in CST Microwave Studio to be numerically simulated. The intensity-dependent metasurface is composed of 6$\times$6 lattices each of which is occupied by 4$\times$4 I-shape meta-atoms. The overall dimensions of the metasurface are 264 mm$\times$264 mm. The 2D and 3D scattering patterns in each case are demonstrated in \textbf{\textcolor{blue}{Figures 6 }}and \textbf{\textcolor{blue}{7}}.
 Under low-power normal plane waves, all lattices behave linearly in identical coding statuses and the metasurface act as an EM mirror regardless of the operating frequency. Therefore, the far-field patterns at 6.7 GHz and 9.3 GHz have single scattered beam pointing at the broadside direction (\textbf{\textcolor{blue}{Figure 7a, c}}). When the metasurface is excited by high-power illuminations, the EM response of nonlinear lattices alters while that of the linear group is kept unchanged. In this case and at the lower frequency band, the nonlinear group of the I-shape meta-atoms traps and dissipates the input signals with more than 99 $\%$ absorption rate while the linear group reflects the input signals with more than 0.86 magnitude. Although the overall response of the metasurface, as the superposition of these distinct responses, is dependent to the nonlinearity distribution over the surface, the far-field patterns of different arrangements show a radar cross section (RCS) reduction of at least 10 dB at the lower frequency band. The corresponding results are given in (\textbf{\textcolor{blue}{Figures 7}}) where a remarkable RCS reduction is obtained in comparison to a metallic reference plane of the same size. At the higher frequency band, the linear and nonlinear lattices treat as the “0” and “1” binary elements with a 180$^\circ$ phase difference, forming a digital metasurface. In such a situation, the nonlinearity distribution becomes very important since different coding sequences of “0” and “1” lattices would lead to diverse types of scattering pattern. Assuming that the reflection phase distribution over the surface is denoted by $\phi_{mn}$, the scattering pattern of the coding metasurface can be expressed as:
\begin{figure*}[t]
	\centering
	\includegraphics[height=2in]{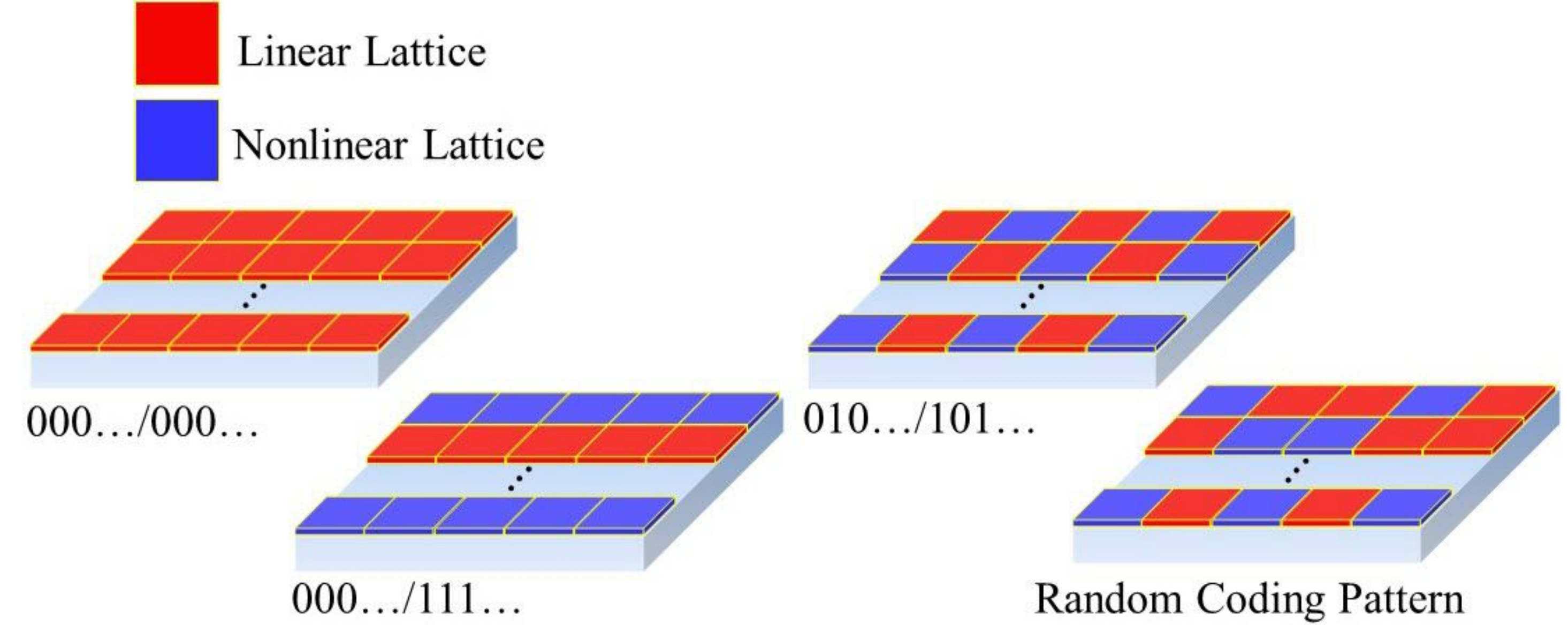}
	\caption{\label{fig:epsart} {Different distributions of nonlinearity over the surface which result in various types of coding pattern seen by a high-power incidence 
			at upper frequency band } }
\end{figure*}
\begin{equation}
F(\theta ,\varphi ) = {f_e}(\theta ,\varphi )\sum\limits_{m = 1}^N {\sum\limits_{n = 1}^N {exp} } \bigg( - i\big\{ {\varphi _{mn}} + kDsin\theta \big[(m - 1/2)cos\varphi  + (n - 1/2)sin\varphi \big]\big\} \bigg)
\end{equation}

in which, $f_e(\theta ,\varphi )$ is the pattern function of the lattice, $\theta$ and $\varphi$  are the elevation and azimuth angles, respectively. $D$ remarks the period of the digital lattices along both x- and y-directions, and $k=2\pi/\lambda$ refers to the wave number in free-space. Based on the generalized Snell's law, endowing the surface with spatially periodic coding sequences yields a metasurface deflecting the incident wave into multiple directions calculated by [7, 18, 21]:

\begin{equation}
{\varphi _1} =  \pm ta{n^{ - 1}}\frac{{{D _x}}}{{{D _y}}},~~ {\varphi _2} = \pi  \pm ta{n^{ - 1}}\frac{{{D _x}}}{{{D _y}}}
\end{equation}
\begin{equation}
\theta  = si{n^{ - 1}}(\lambda \sqrt {\frac{1}{{D _x^2}} + \frac{1}{{D _y^2}}} )
\end{equation}
\begin{figure*}[t]%
	\centering
	\subfigure[]{%
		\label{fig:21}%
		\includegraphics[height=1.4in]{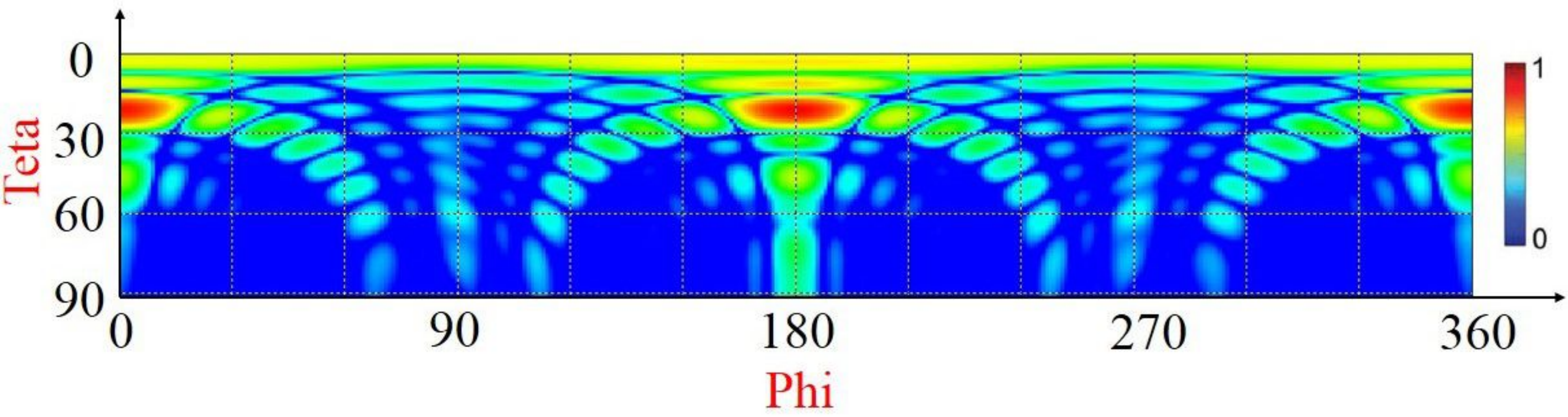}}%
	
	\subfigure[][]{%
		\label{fig:22}%
		\includegraphics[height=1.4in]{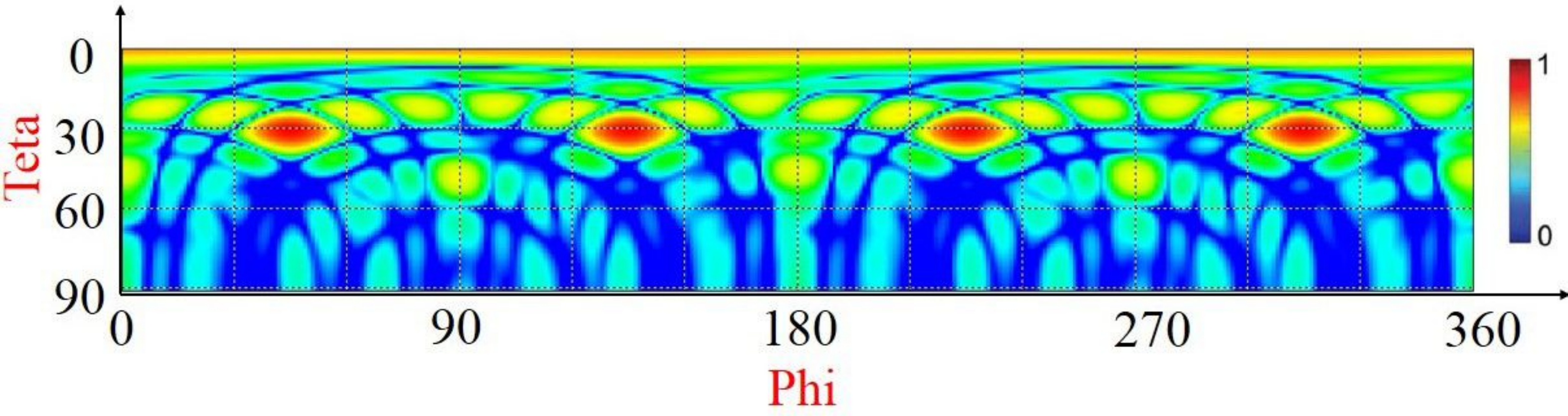}}%
	\caption{2-D far-field results that show the performance of the metasurface upon illuminating by high-power incidences for (a) 000..0/11...1 coding sequence along the x-direction and (b) 01..0/10...1 coding sequence along both x- and y-directions. }
	\label{fig:4}
\end{figure*}

\begin{figure*}[t]
	\centering
	\includegraphics[height=3.1in]{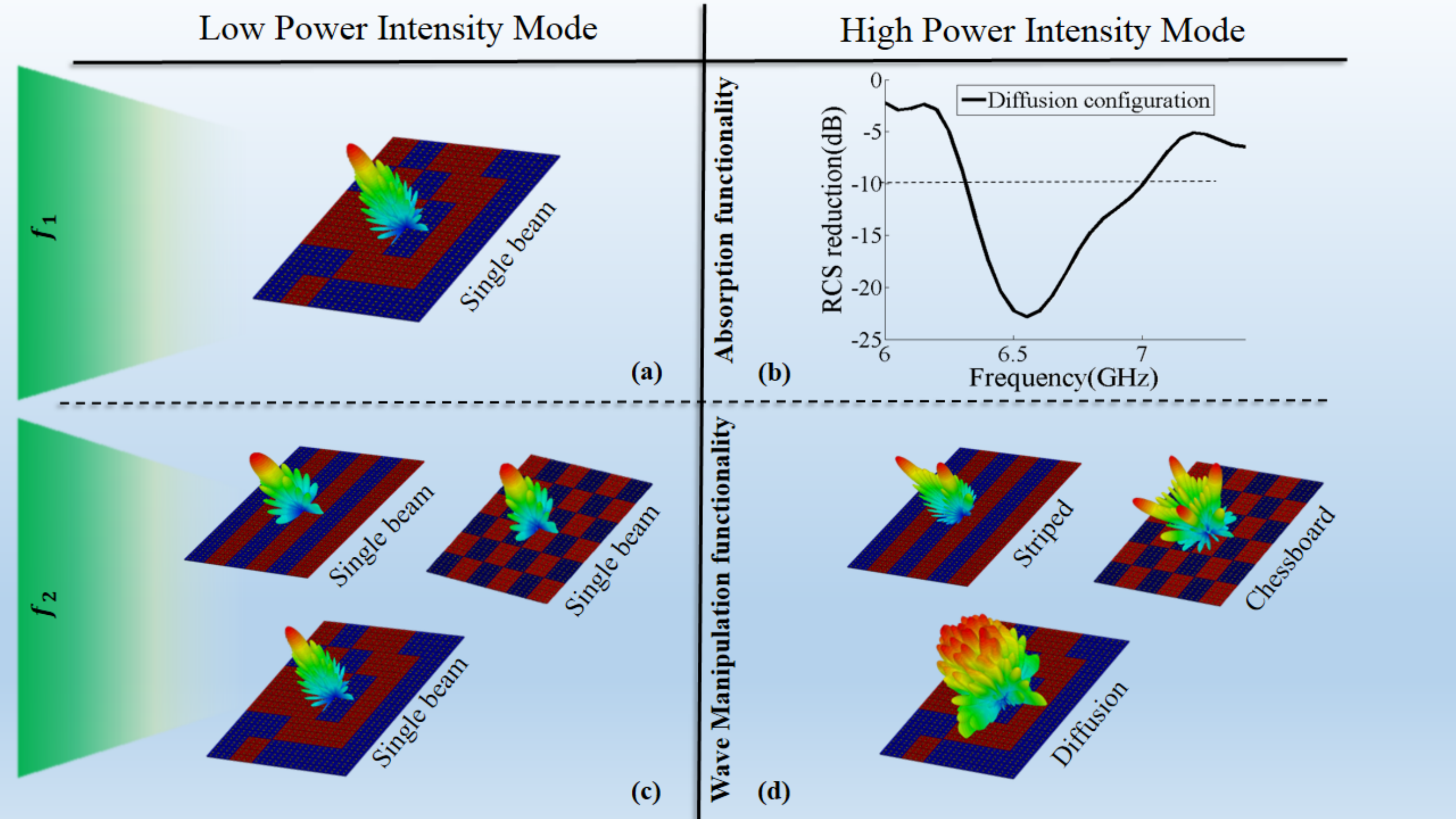}
	\caption{\label{fig:epsart} {:The 3D far-field patterns of the designed metasurface for (a), (c) low and (b), (d) high illumination powers at the (a), (b) lower and (c), (d) upper frequency bands. } }
\end{figure*}

wherein, $D_x$ and $D_y$ indicate the spatial periods of the coding sequence along the horizontal and vertical directions, respectively. For instance, being programmed with the periodic coding sequence of 010101.../010101..., the nonlinear metasurface will split the normally incident wave into two symmetrically oriented beams along $(\theta_1,\phi_1)$=(11$^\circ$,0$^\circ$) and $(\theta_2,\phi_2)$=(11$^\circ$,180$^\circ$), whereas under the periodic coding sequence of 010101.../101010..., the incident beam will  be mainly reflected to four symmetrically oriented directions $(\theta_1,\phi_1)$=(15$^\circ$,45$^\circ$), $(\theta_2,\phi_2)$=(15$^\circ$,135$^\circ$), $(\theta_3,\phi_3)$=(15$^\circ$,225$^\circ$), and $(\theta_4,\phi_4)$=(15$^\circ$,315$^\circ$). The numerical results illustrated in \textbf{\textcolor{blue}{Figures 6a, b}} and \textbf{\textcolor{blue}{Figure 7d}} confirm well our theoretical predictions for the tilt angles. 

Lower scattering signatures can be achieved through uniform re-distribution of the reflected beams at the upper half-space, yielding a speckle-like far-field pattern. The entropy-based method described in \cite{rajabalipanah2019ultrabroadband,momeni2018information} has been utilized here to seek for the optimum coding pattern resulting in a highly-efficient scattering diffusion. This approach is more efficient and faster than its time-consuming counterparts which search between a large space including all $2^{N^2}$ possible solutions \cite{rajabalipanah2019ultrabroadband,momeni2018information}. Entropy is a key measure to interpret the far-field patterns generated by a digital metasurface with randomized distribution of “0” and “1” elements.  In order to ensure the maximum uniformity of the scattering pattern, the entropy level describing the randomness of the coding pattern should be maximized \cite{rajabalipanah2019ultrabroadband,momeni2018information}. The optimization has been carried out by using the Bat algorithm in the same manner given in \cite{momeni2018information} to find a fully-randomized coding pattern via searching between only N-bits binary sequences. The maximum entropy level has been attained as high as 1.74 while that of the reference plane and the chessboard arrangement is about 0.135 and 0.442, respectively. \textbf{\textcolor{blue}{Figure 7d}} shows the 3D far-field pattern of the randomized coding metasurface at 9.39 GHz, where the high-power incident beam is scattered into numerous beams oriented along random directions. This remarks that the both monostatic and bistatic RCS signatures of the nonlinear metasurface at higher band are dramatically reduced upon illuminating by high-power input signals	, as if the metasurface plays the role of an EM fuse or a high-power protection interface. . The numerical simulations verify the tri-state performance of the proposed nonlinear metasurface, i.e. mirroring for the low-power incidences, and energy harvesting and scattering diffusion for the high-power illuminations of 6.7 GHz and 9.39 GHz, respectively. In fact, the nonlinear metasurface smartly acts as a bi-spectral low-scattering surface upon high-power threats of 6.7 GHz and 9.39 GHz with exposing a low RCS level (\textbf{\textcolor{blue}{Figure 7}}).

\section{Conclusion}
~~In summary, we proposed a tri-state bi-spectral power-dependent metasurface with arbitrary distribution of nonlinearity over the surface. The designed metasurface is composed of several diode-loaded I-shape meta-atoms whose EM responses are determined by the intensity of the incident fields without need to any active biasing circuit. Being excited by low-power radiations, the nonlinear architecture plays the role of an EM mirror while under high-power illuminations, it acts as an energy harvesting platform and an EM diffuser at the lower and higher frequency bands, respectively. It was shown that different distribution of nonlinearity over the surface would lead to diverse types of the scattering patterns at the higher frequency band. The numerical simulations verify the tri-state performance of the proposed nonlinear metasurface. 

\bibliography{sample}






\end{document}